# LOCALIZATION OF CLASSICAL WAVES IN WEAKLY SCATTERING TWO-DIMENSIONAL MEDIA WITH ANISOTROPIC DISORDER

**Gregory Samelsohn[1,2] and Valentin Freilikher[2,3]**

[1] Department of Electrical and Electronic Engineering,
Negev Academic College of Engineering, Beer-Sheva 84100, Israel

[2] The Jack and Pearl Resnick Institute of Advanced Technology, Department of Physics,
Bar-Ilan University, Ramat-Gan 52900, Israel

[3] Complex Photonic Systems, Department of Science and Technology,
University of Twente, P.O. Box 217, 7500 AE Enschede, The Netherlands

**Abstract**

We study the localization of classical waves in weakly scattering 2D systems with anisotropic disorder. The analysis is based on a perturbative path-integral technique combined with a spectral filtering that accounts for the first-order Bragg scattering only. It is shown that in the long-wavelength limit the radiation is always localized, and the localization length is independent of the direction of propagation, the latter in contrast to the predictions based on an anisotropic tight-binding model. For shorter wavelengths that are comparable to the correlation scales of the disorder, the transport properties of disordered media are essentially different in the directions along and across the correlation ellipse. There exists a frequency-dependent critical value of the anisotropy parameter, below which waves are localized at all angles of propagation. Above this critical value, the radiation is localized only within some angular sectors centered at the short axis of the correlation ellipse and is extended in other directions.

Anderson-type localization of classical waves in disordered systems is a topic of increasing current interest due to its fundamental role in wave-matter interactions, and also due to the significance of possible applications [1]. While the localization in one-dimensional (1D) random systems has been studied in considerable detail, a quantitative analytical description of the phenomenon in higher dimensions still presents a challenge. The question here concerns the relationship between the localization and its characteristics (say, localization length), on the one hand, and the correlation properties of the scattering potential, on the other. However, most of the existing results are related to a delta-correlated potential or obtained numerically by using discrete schemes, such as the tight-binding model, and therefore are relevant mainly for low-frequency excitations. Moreover, the theories describing wave localization have been developed primarily for systems with isotropic disorder, and those considering anisotropy [2,3], are not directly applicable to classical waves. The only exception is the limiting case of an infinite correlation scale in one direction (randomly stratified media), which has been studied rather comprehensively [4]. In such media, the radiation is localized in the direction across the layers and is typically channeled along the layers, similar to that occurring in a regular waveguide. Although this model can be useful for understanding the basic mechanisms of wave localization in anisotropic systems, it is of little help when media with finite anisotropy are concerned. Experimental studies dealing with both electronic [5] and classical wave [6,7] transport in anisotropic systems have been initiated only recently and are far from being complete.

The present study is based on a path integral approach [8] that enables a perturbative analysis of the localization length in random media with continuous type disorder described by an arbitrary correlation function. It is shown that in the long-wavelength limit, a two-dimensional (2D) anisotropic system is characterized by a finite localization length, which is independent of the direction of propagation. For shorter wavelengths, comparable to the correlation scale of the disorder, the calculations demonstrate that waves should be exponentially localized within some angular sectors centered at the short axis of the correlation ellipse, and are extended in other directions. The crossover from isotropic to anisotropic media resembles a phase transition: for any given wave number there is a frequency-dependent critical degree of anisotropy below which the localization length is finite for all angles of propagation. Above this critical value, one observes channeling of the wave energy along the physically predetermined directions.

We consider the radiation of a point source located at the origin in an infinite statistically homogeneous medium which is characterized by the relative permittivity distribution $\varepsilon(\mathbf{R}) = 1 + \tilde{\varepsilon}(\mathbf{R})$. The "scattering potential" $\tilde{\varepsilon}(\mathbf{R})$ is assumed to be a zero-mean random perturbation with a given correlation function, $B_\varepsilon(\mathbf{R})$, whose Fourier transformation defines the power spectrum, $\Phi_\varepsilon(\mathbf{K})$. Thus, at any point $\mathbf{R}$, the time-harmonic wave field satisfies the Helmholtz equation,

$$\nabla^2 G(\mathbf{R}) + k^2 [1 + \tilde{\varepsilon}(\mathbf{R})] G(\mathbf{R}) = -\delta(\mathbf{R}) , \tag{1}$$

where $k = 2\pi/\lambda$ is the wave number in an unperturbed homogeneous medium.

The objective is to find the self-averaging (non-random) value called inverse localization length (Lyapunov exponent), which is defined as

$$\xi^{-1}(\mathbf{k}) = -\lim_{L\to\infty} L^{-1} \ln I(L, \mathbf{k}) , \tag{2}$$

where $I(L, \mathbf{k}) \equiv |G(\mathbf{R})|^2$ is the wave intensity measured at a distance $L \equiv |\mathbf{R}|$ from the point source, and $\mathbf{k} = k\,\mathbf{R}/R$ is the wave vector directed along the line connecting the source with the observation point. The positiveness of $\xi^{-1}(\mathbf{k})$ means that the wave intensity in typical realizations decays exponentially, $I(L, \mathbf{k}) \sim \exp[-L/\xi(\mathbf{k})]$, i.e., the radiation is localized in the $\mathbf{k}$-direction. To calculate the Lyapunov exponent, we use the following indirect procedure consisting of two steps. We first evaluate the decrement of the mean intensity,

$$\alpha(\mathbf{k}) = -\lim_{L\to\infty} L^{-1} \ln \langle I(L, \mathbf{k}) \rangle , \tag{3}$$

which, in general, does not coincide with the inverse localization length, because, being obtained by averaging over the ensemble of all possible realizations of the scattering potential, it is formed mainly by the contribution of low probable realizations with high-$Q$ resonances [9]. However, by evaluating the decrement $\alpha(\mathbf{k})$ in the second order of the scattering potential and presenting the result in the form of an integral expansion in the $\mathbf{K}$-space [see Eq. (6) below], we can filter out the contribution of low probable resonances, and therefore estimate the inverse localization length, the second essential step in our calculations.

To realize this program, we use the method of proper time [10], according to which the solution of Eq. (1) is given in the form

$$G(\mathbf{R}) = \frac{i}{2k} \int_0^\infty d\tau \exp(\, i k\tau/2\,)\, g(\mathbf{R}, \tau) , \tag{4}$$

where function $g(\mathbf{R},\tau)$ satisfies a Schrödinger like equation,

$$2ik\partial_\tau g + \nabla^2 g + k^2\tilde{\varepsilon}(\mathbf{R})\, g(\mathbf{R},\tau) = 0 \ , \qquad \tau > 0 \ , \tag{5}$$

supplemented by an appropriate initial condition, $g(\mathbf{R},0) = \delta(\mathbf{R})$. By applying the path integral solution to the propagator $g(\mathbf{R},\tau)$, the mean intensity of the wave is constructed in the form allowing for the averaging to be easily performed. The $\tau$ integrals are then evaluated by the method of stationary phase, and the double path integral is replaced with a first cumulant approximation, which in accordance with Eq. (3) yields [11]:

$$\alpha(\mathbf{k}) = \frac{\pi}{2}k^3 \int d\mathbf{K}\, f(\mathbf{k},\mathbf{K})\Phi_\varepsilon(\mathbf{K}) \ . \tag{6}$$

Here,

$$f(\mathbf{k},\mathbf{K}) = K^{-1}\,\delta(K - |2\mathbf{k}\cdot\mathbf{K}/K|) - K^{-2}\,\vartheta(K - |2\mathbf{k}\cdot\mathbf{K}/K|) \ , \tag{7}$$

$\delta(x)$ is the Dirac $\delta$ function, and $\vartheta(x)$ is the Heaviside step function. It is important to note that Eqs. (6) and (7) are valid for any power spectrum $\Phi_\varepsilon(\mathbf{K})$ and have the same general form for random systems of any dimensionality. In deriving Eq. (6), we have assumed that the distance $L$ is much larger than any correlation scale of the disorder. This does not allow considering the crossover to a randomly layered medium, where one of the correlation scales is obviously infinite. However, this model is suitable for studying the transition from isotropic random media to quasi-layered structures in which one of the correlation scales is much greater than other(s).

To extract the value of $\xi^{-1}(\mathbf{k})$ from Eq. (6), we have to analyze first the basic mechanisms determining the interaction of the wave field with different spectral components of the scattering potential. To this end, let us recall the well known perturbative result for the inverse localization length in 1D systems [9]:

$$\xi^{-1}(k) = \frac{\pi}{2}k^2\Phi_\varepsilon(2k) \ . \tag{8}$$

According to this expression, only $K = \pm 2k$ components of the spectrum contribute to $\xi^{-1}(k)$ for a time-harmonic wave with wave number $k$, which means that the localization is caused by the Bragg scattering on the corresponding periodic lattice. As can be easily verified, Eq. (8) is reproduced exactly, if we keep only $\pm 2k$ Bragg components in Eq. (6) written for the 1D case. High frequency tail, $K > 2k$, of the power spectrum does not contribute to the localization length in weakly scattering media, but is related to the spikes of wave intensity and enhanced transmission in resonant realizations [12].

It is natural to extend this algorithm to higher dimensional (2D and 3D) media, where, according to the macroscopic point of view [13], the localization appears as a result of a subtle interplay between different periodic lattices, which constitute any realization of a random medium. As is known, in multidimensional systems the components of the power spectrum (vectors **K** of the reciprocal lattices) that participate in the Bragg scattering and, therefore, contribute to the localization length, are located within the limiting circle (2D) or sphere (3D) of radius $2k$ in the Ewald construction, see Fig. 1 (in 1D, this diagram degenerates into three points, $K = 0, \pm 2k$, only) [8]. Hence, to estimate the value of $\xi^{-1}(\mathbf{k})$ in higher dimensions in the same way as it was done in the 1D case, we should perform spectral filtering by reducing the integration domain in Eq. (6) to the limiting sphere, $K \le 2k$, that results in

$$\xi^{-1}(\mathbf{k}) = \frac{\pi}{2} k^3 \int\limits_{K \le 2k} d\mathbf{K}\; f(\mathbf{k},\mathbf{K}) \Phi_\varepsilon(\mathbf{K}) \,. \tag{9}$$

The kernel of this integral transform, function $f(\mathbf{k},\mathbf{K})$ given by Eq. (7), consists of two terms having opposite signs, which means that in many-dimensional systems different spectral components of the scattering potential control in radically different ways the propagation of wave at typical realizations. While the localization is favoured by interaction of the wave with spectral components lying on the curve $K = |2\mathbf{k} \cdot \mathbf{K}/K|$, wave scattering on lattices bounded by this curve and the limiting circle (see Fig. 1) suppresses localization and can even destroy it. Thus, the result of the competition between these two contributions and the exact answer to the question whether the wave is localized or not is determined ultimately by the structure of the power spectrum.

In what follows we consider two-dimensional random systems. If the disorder is isotropic, the integration over angular coordinate in Eq. (9) can be performed analytically [8]. In general case of an angle-dependent power spectrum $\Phi_\varepsilon(K,\phi')$, Eq. (9) can be presented in the form

$$\xi^{-1}(\mathbf{k}) = \pi k^3 \int_0^\pi d\phi' \left[ \Phi_\varepsilon(2k\beta,\phi') - \int_\beta^1 dx\, x^{-1} \Phi_\varepsilon(2kx,\phi') \right] , \tag{10}$$

where $\beta = \left|\cos(\phi - \phi')\right|$, and $\phi$ is the angle of propagation, so that $\mathbf{k} \equiv (k, \phi)$. Note, that the interval of integration over $\phi'$ is reduced to $[0, \pi]$ since $\beta(\phi' + \pi) = \beta(\phi')$, and the power spectrum is a periodic function in polar coordinates, $\Phi_\varepsilon(K, \phi' + \pi) = \Phi_\varepsilon(K, \phi')$.

To exemplify the result obtained, we have assumed that the medium is described by an anisotropic Gaussian correlation function,

$$B_\varepsilon(\mathbf{R}) \equiv B_\varepsilon(x, y) = \sigma_\varepsilon^2 \exp\left(-\mu x^2/l_\varepsilon^2 - y^2/\mu l_\varepsilon^2\right), \tag{11}$$

where $\sigma_\varepsilon^2$ is the variance of the fluctuations, while $l_\varepsilon$ and $\mu$ are, respectively, the mean geometrical value and the ratio of the correlation lengths along the two coordinate axis. When, for instance, $\mu > 1$, the inhomogeneities are stretched along the *y*-axis. Since the two situations, $\mu > 1$ and $\mu < 1$, are topologically equivalent, we consider only the case $\mu > 1$, which means that $\phi$ is the angle measured between the short axis of the correlation ellipse and the direction of wave propagation. The power spectrum corresponding to Eq. (11) is given by

$$\Phi_\varepsilon(K, \phi') = (1/4\pi)\,\sigma_\varepsilon^2\, l_\varepsilon^2 \exp\left(-a_\mu^2\, l_\varepsilon^2 K^2/4\right), \tag{12}$$

where

$$a_\mu^2 \equiv a_\mu^2(\phi') = \mu \sin^2\phi' + (1/\mu)\cos^2\phi' \ . \tag{13}$$

By substituting the latter expressions into Eq. (10) we obtain

$$\xi^{-1}(\mathbf{k}) = \frac{1}{8}\sigma_\varepsilon^2 \kappa^3 l_\varepsilon^{-1} \int_0^\pi d\phi' \left[2\exp\left(-a_\mu^2\kappa^2\beta^2\right) + E_1\left(a_\mu^2\kappa^2\right) - E_1\left(a_\mu^2\kappa^2\beta^2\right)\right], \tag{14}$$

where $E_1(x)$ is the exponential integral, and $\kappa = kl_\varepsilon$ is the normalized wave number.

The inverse localization length in an isotropic system ($\mu = 1$) as a function of $\kappa$ is shown in Fig. 2. Although $\xi^{-1}(k)$ has a well-defined maximum at some intermediate frequency band, in the high-frequency limit the localization length is independent of the wavelength. The same effect has been observed recently in numerical simulations [14] where the localization length for 2D strongly disordered systems was shown to saturate at high frequencies. While in Ref. [14] the saturation was attributed to the discontinuous character of the permittivity distribution, our consideration shows that this effect can be of universal nature.

For anisotropic systems, in the long wavelength limit, $\mu\kappa^2 << 1$, Eq. (14) takes the form

$$\xi^{-1}(k) = (\pi/4)(1 - \ln 2)\,\sigma_\varepsilon^2 \kappa^3 l_\varepsilon^{-1} \ , \tag{15}$$

which means that $\xi^{-1}(k)$ does not depend on the angle of propagation under specified conditions, and the wave is localized in all directions as it is in isotropic media.

Anisotropy of the system shows up when the radiation wavelength becomes comparable to the correlation scale of the disorder, see Fig. 3. While at $\kappa = 1/4$ [Fig. 3(a)], radiation is still localized in all directions for all values of $\mu$ used in our calculations, for shorter wavelengths, already for $\kappa = 1/2$ [Fig. 3(b)], high degree of anisotropy causes the loss of complete localization: at some critical $\mu$-dependent angle, $\phi_c$, the inverse localization length turns to zero and there appear angular sectors centered at the long axis of the correlation ellipse, within which the wave is extended. For $\kappa = 1$ [Fig. 3(c)], only weakly anisotropic media ($\mu \leq 2$) could localize waves in all directions. For increasing values of $\kappa$, even very moderate degree of anisotropy destroys complete localization. Moreover, as can be seen from the plots corresponding to $\kappa = 2$ [Fig. 3(d)] and, especially, $\kappa = 4$ [Fig. 3(e)], the dependence of the localization length on the angle of propagation becomes more complicated: a unimodal angular structure typical of small $\kappa$ values, transforms into a bimodal distribution. In other words, the maximum of the inverse localization length (the strongest localization) is observed not for waves propagating in the transverse direction, i.e., along the $\phi = 0$ axis, as one could expect, but at some intermediate angle ($0 < \phi < \phi_c$), that depends on both $\kappa$ and $\mu$.

This, at first glance counterintuitive, behavior of the localization length can be understood if we remind that the same physical mechanism, namely, scattering on the resonant Bragg lattices, underlies not only band gap formation in periodic systems, but also wave localization in random media. There exist a number of examples, in particular, in two dimensions, which show a complete stop band only above some critical value of the refractive index contrast between the constituent dielectrics forming a photonic crystal. For smaller contrasts, the wave can freely escape the photonic trap, being channeled within some angular sectors defined by the structure [15]. Since there is a parallel between the bandgap phenomena in photonic crystals, on the one hand, and our treatment of classical wave localization in random media, on the other, it is of little surprise that the anisotropy in correlations of the scattering potential leads to similar effects. Also, one can find an interesting analogy to the angle resolved picture of wave transport through 3D disordered photonic crystals, where at some frequencies a bimodal angular distribution of the radiation has been observed [16].

In summary, the localization of classical waves in 2D random media with anisotropic disorder has been considered. The physical analysis is based on the formula that generalizes the well known 1D Bragg scheme to multidimensional systems. It is shown that in contrast to the low-frequency regime where the radiation is localized in all directions, for shorter wavelengths wave transport becomes highly anisotropic. The important prediction is that complete localization may be absent in 2D anisotropic systems when the wavelength is comparable to the correlation scales of the disorder. A similar analysis based on Eq. (9) can be performed also for 3D anisotropic media. The predicted effect of anisotropy on wave transport in random media may find applications in photonics and other related technologies. For instance, in many cases the degree of anisotropy can be controlled much easier than the refractive index contrast [17]. The dependence of $\xi^{-1}(\mathbf{k})$ on the anisotropy open, therefore, a way to control localization in disordered systems. In particular, such effects may be used in random lasers [18] where the possibility of changing the shape of lasing modes and, hence, their quality factor and lasing threshold, add considerably to the technique.

**Acknowledgment**

This work was supported by the Israeli Science Foundation (grant 328/02).

## FIGURE CAPTIONS

**Fig. 1.** Momentum diagram, representing schematically the process of Bragg scattering in a weakly disordered medium. The points of the Ewald circle determine all possible spectral components (vectors $\mathbf{K} = \mathbf{k}' - \mathbf{k}$ of the power spectrum) that could transform the incident wave ($\mathbf{k}$) into a scattered one ($\mathbf{k}'$). The limiting circle bounds all spectral components coupling any two wave vectors in the process of elastic scattering.

**Fig. 2.** Inverse localization length $\xi^{-1}(\mathbf{k})$ as a function of the normalized wave number $\kappa = kl_\varepsilon$ for 2D statistically isotropic medium with Gaussian correlation function. The localization length is normalized to $\sigma_\varepsilon^2$ and is given in units of $l_\varepsilon$.

**Fig. 3.** Inverse localization length $\xi^{-1}(\mathbf{k})$ as a function of the angle of propagation $\phi$ for 2D statistically anisotropic media, with different values of the anisotropy parameter: $\mu = 2$ (dotted line), $\mu = 4$ (dash-dotted line), $\mu = 8$ (dashed line), $\mu = 16$ (solid line). Straight dashed line corresponds to the isotropic case. The localization length is normalized to $\sigma_\varepsilon^2$ and is given in units of $l_\varepsilon$. Each of the five plots corresponds to a given value of the normalized wave number $\kappa = kl_\varepsilon$: (a) $\kappa = 1/4$; (b) $\kappa = 1/2$; (c) $\kappa = 1$; (d) $\kappa = 2$; (e) $\kappa = 4$.

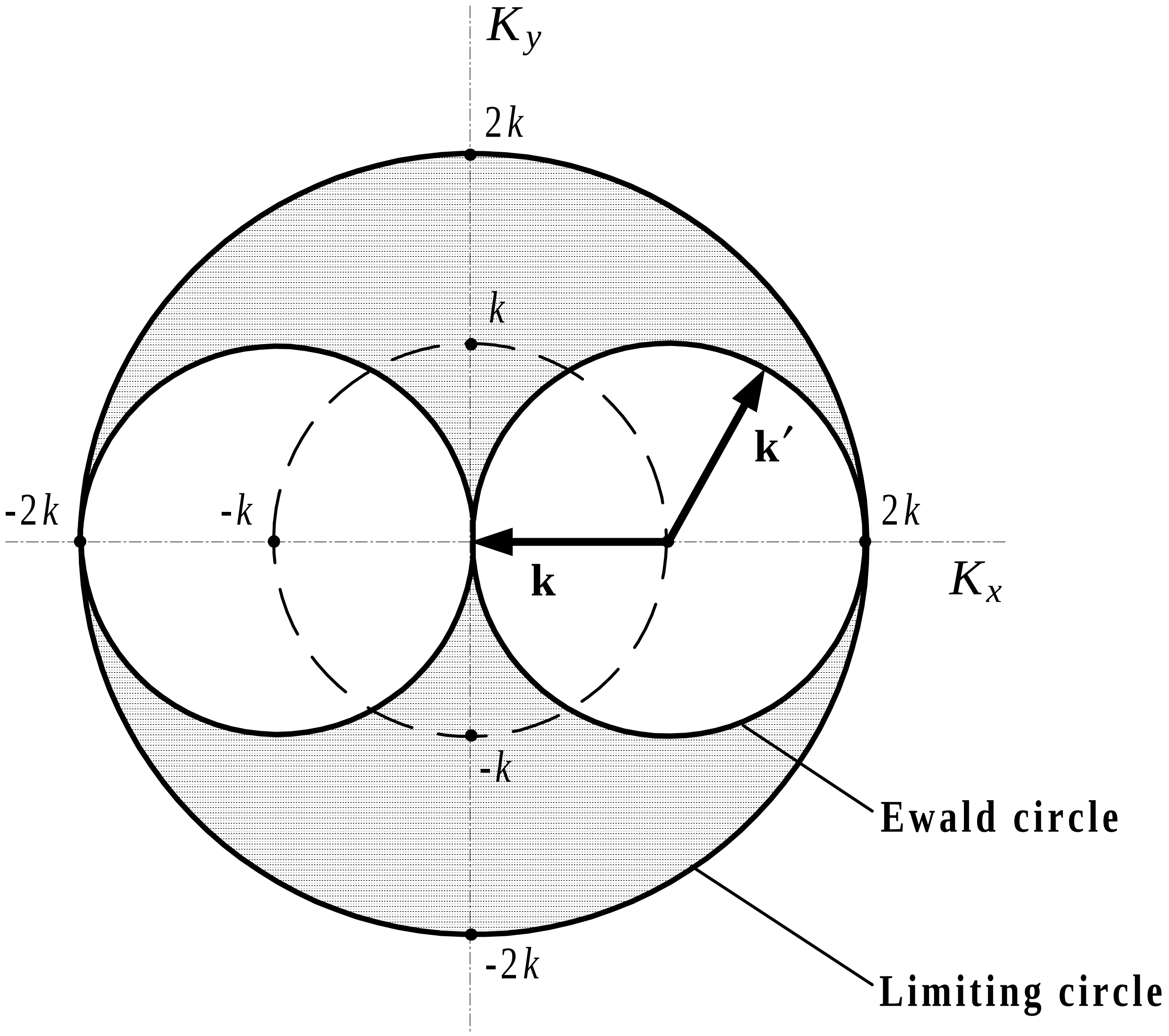

Fig. 1.

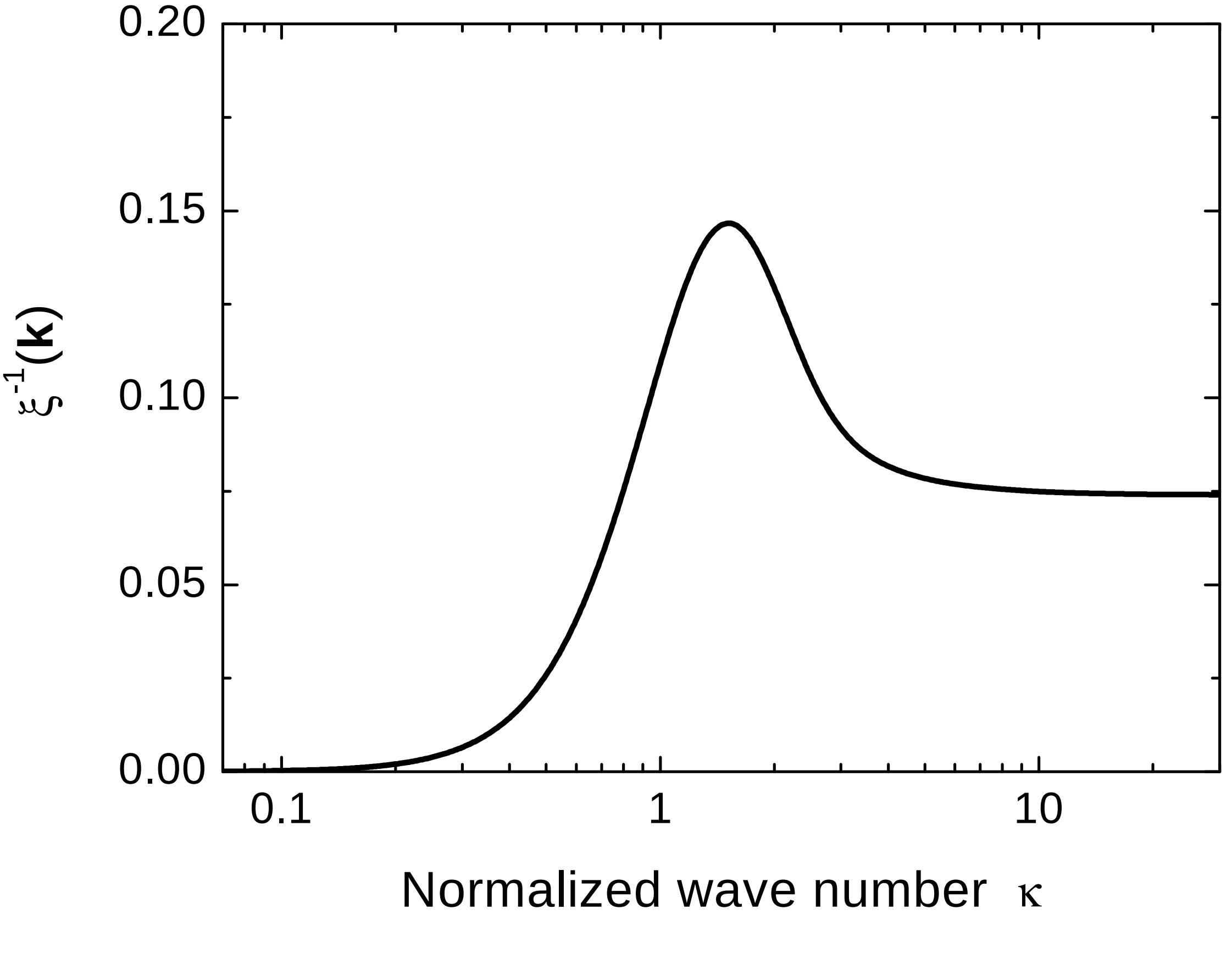

Fig. 2.

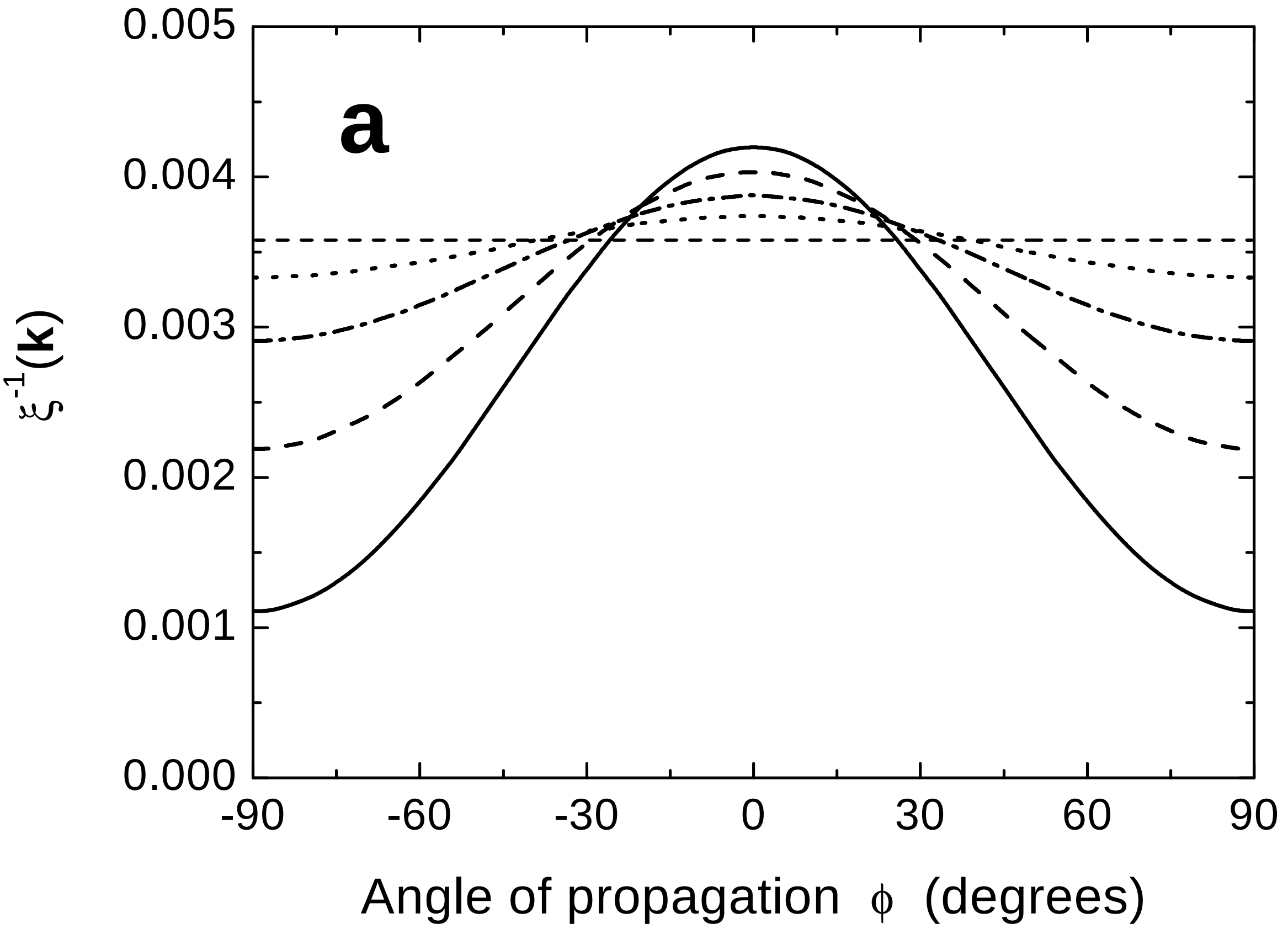

Fig. 3a.

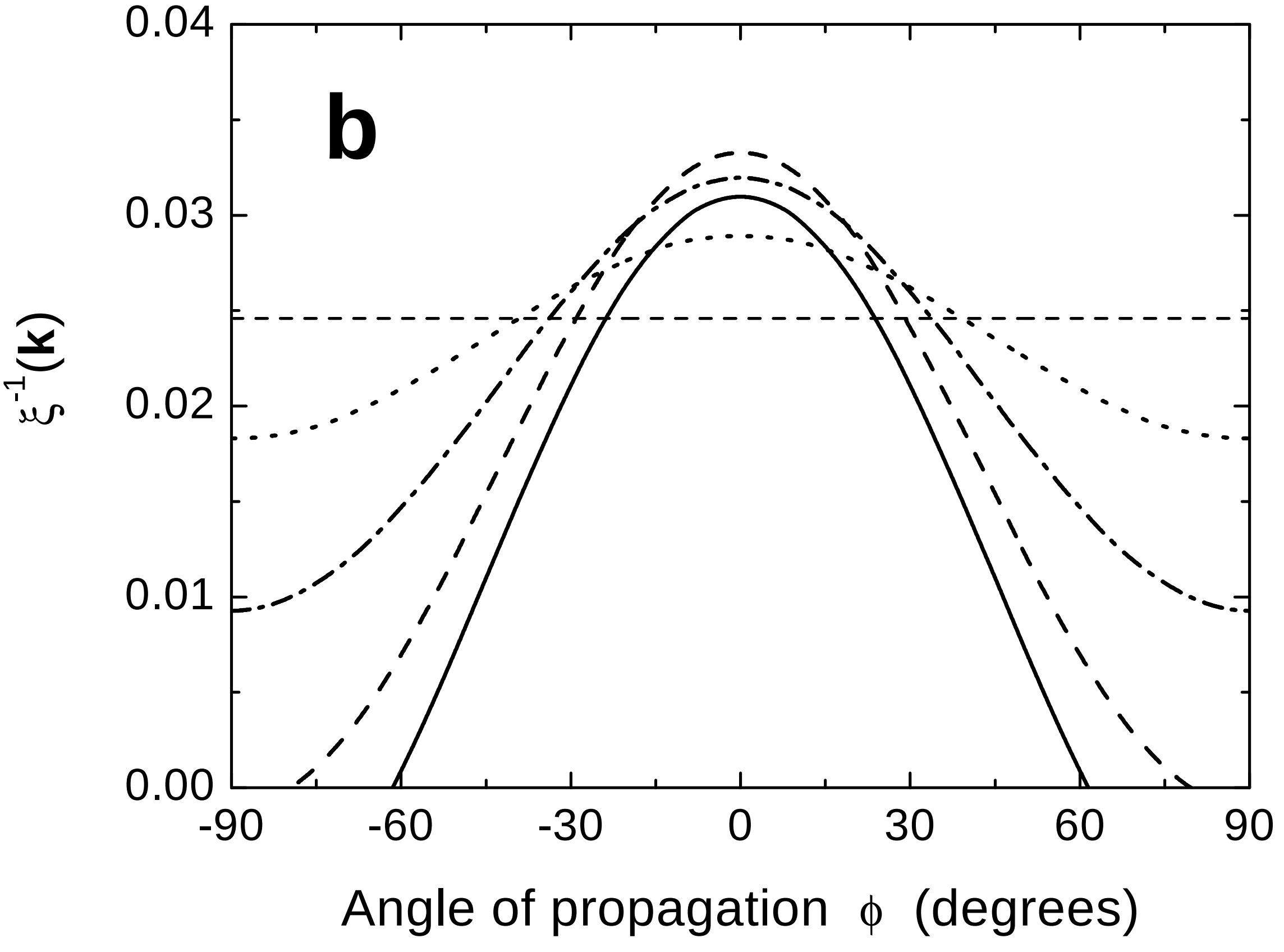

Fig. 3b.

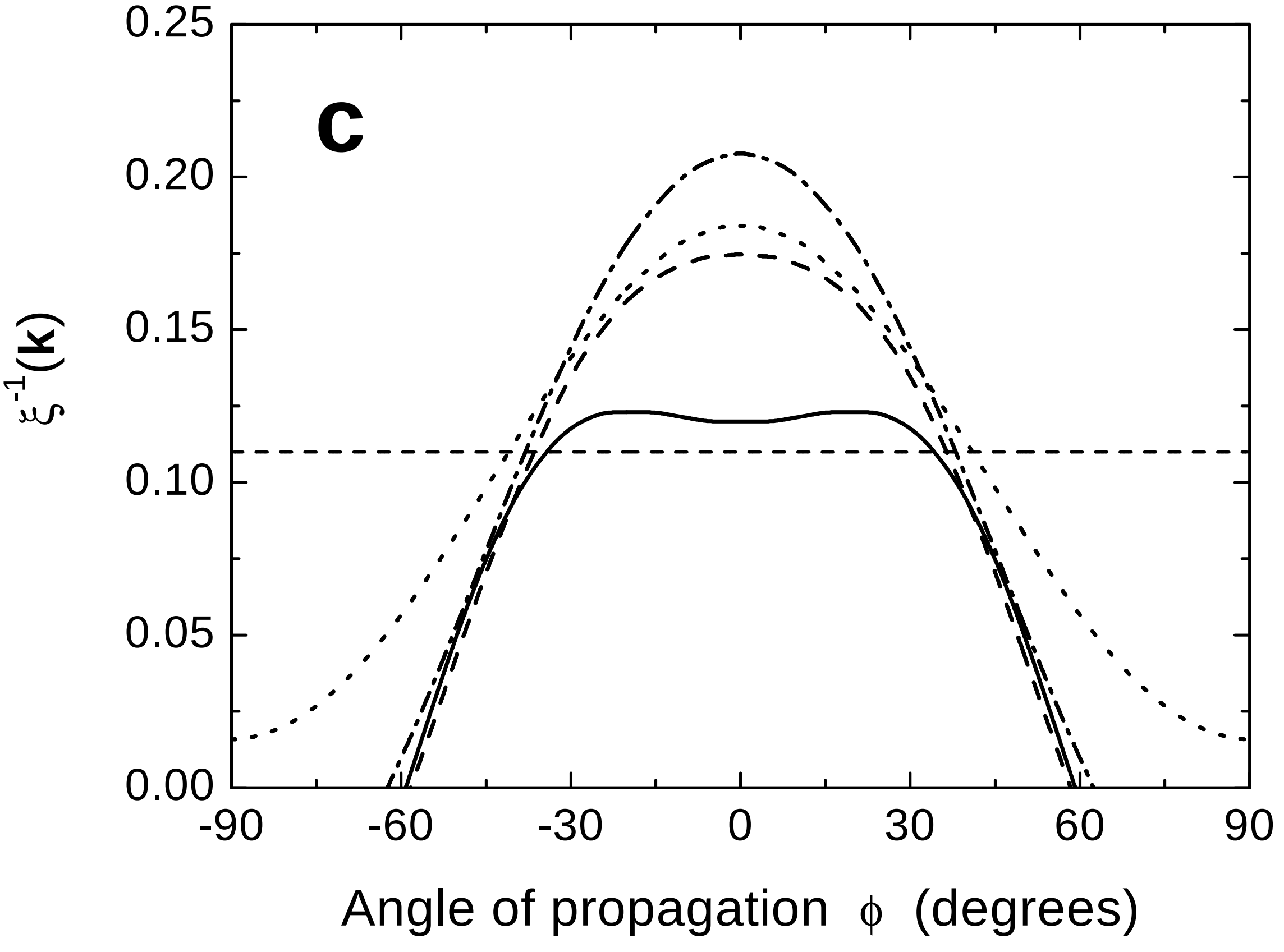

Fig. 3c.

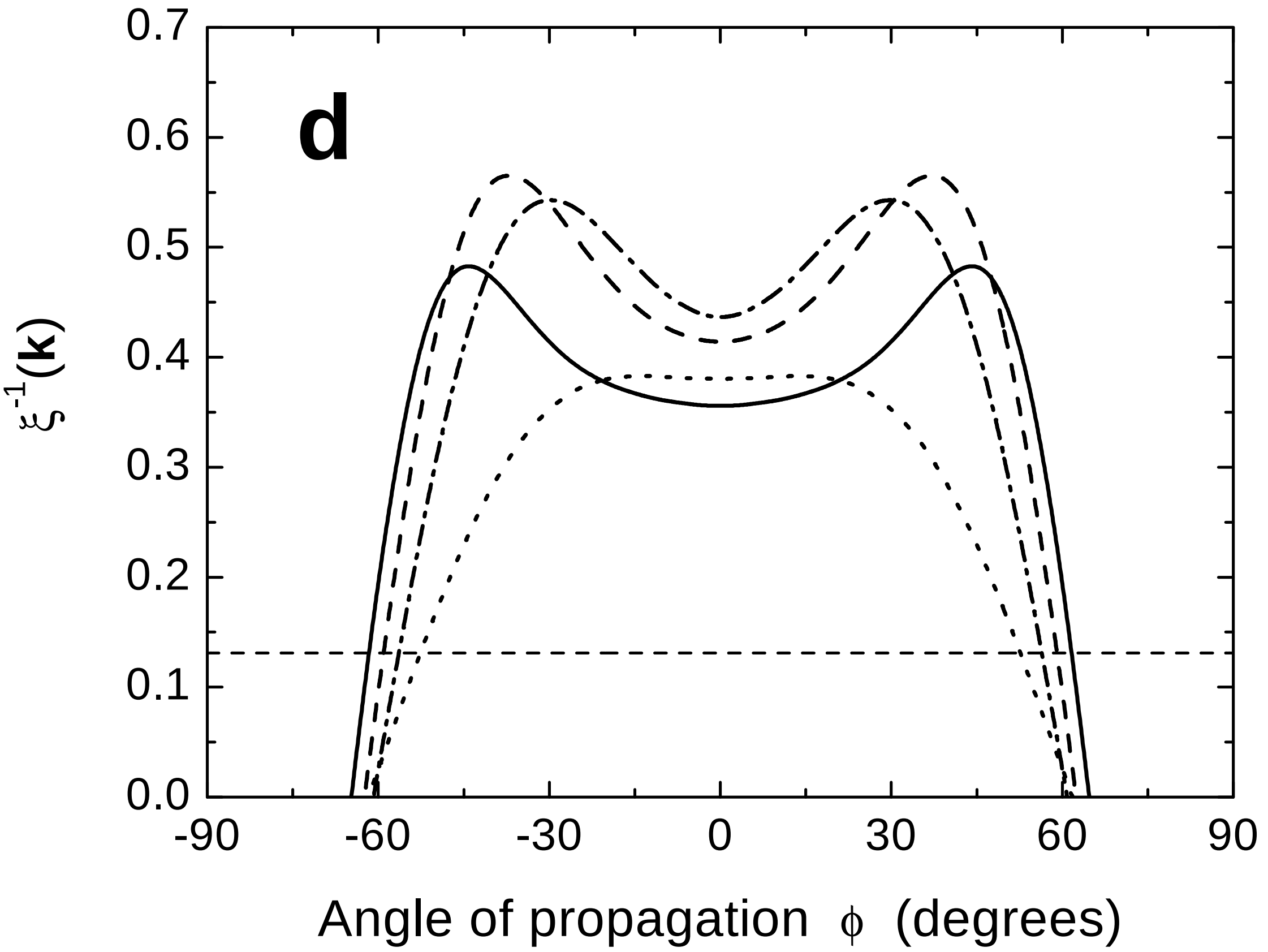

Fig. 3d.

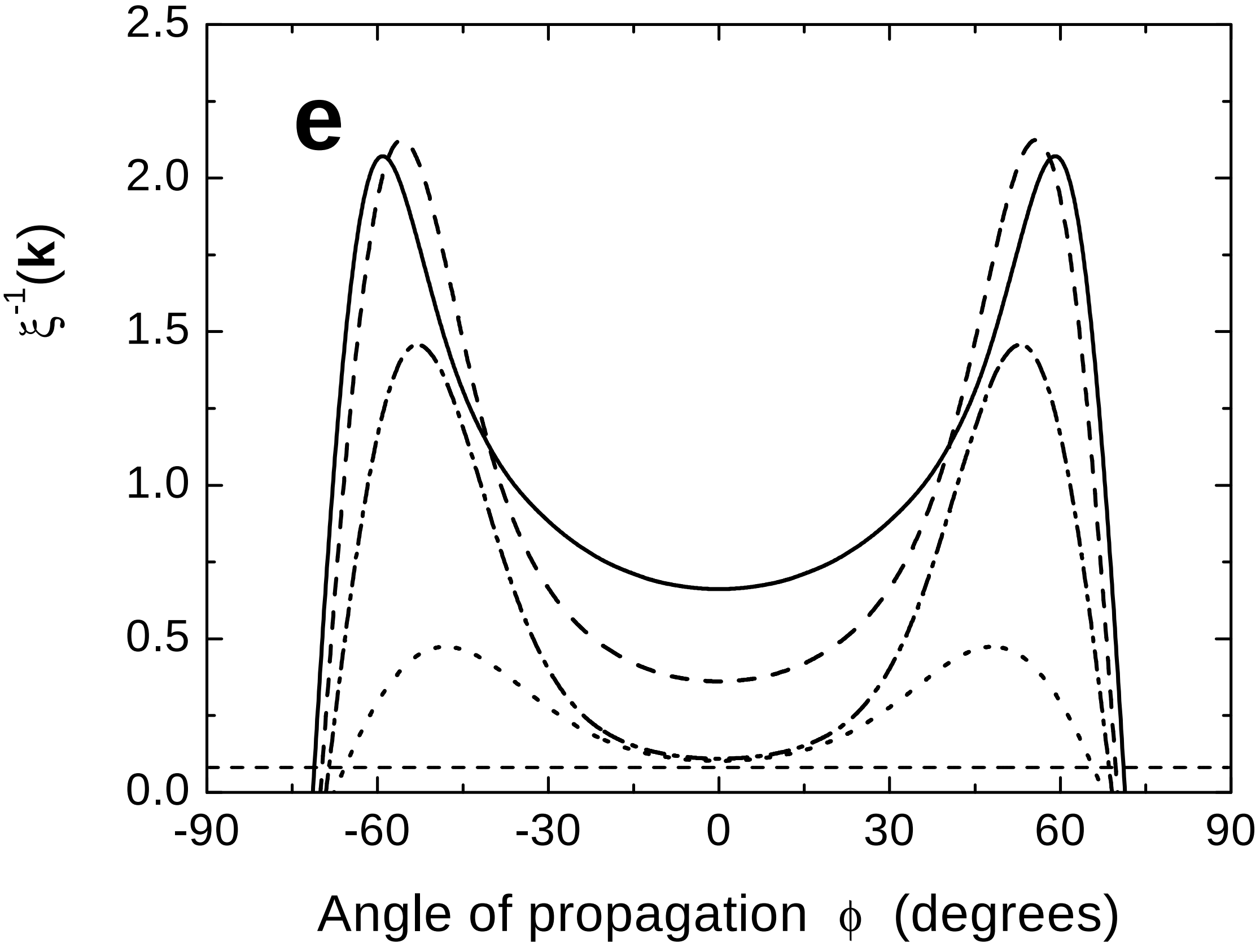

Fig. 3e.